# Revisiting greenhouse gases adsorption in carbon nanostructures: advances through a combined first-principles and molecular simulation approach

Henrique M. Cezar[1,2], Teresa D. Lanna[1], Daniela A. Damasceno[1], Alexsandro Kirch[1], Caetano R. Miranda[1]

[1]Universidade de Sao Paulo, Instituto de Fisica, Rua do Matao 1371, Sao Paulo, SP 05508-090, Brazil

[2]Hylleraas Centre for Quantum Molecular Sciences and Department of Chemistry, University of Oslo, PO Box 1033 Blindern, 0315 Oslo, Norway

Corresponding author: Caetano R. Miranda, email: crmiranda@usp.br

**Abstract:**

Carbon nanostructures are promising materials to improve the performance of current gas separation membrane technologies. From the molecular modeling perspective, an accurate description of the interfacial interactions is mandatory to understand the gas selectivity in the context of greenhouse gases applications. Most of the molecular dynamics simulations studies considered available force fields with the standard Lorentz-Berthelot (LB) mixing rules to describe the interaction among carbon dioxide ($CO_2$), methane ($CH_4$) and carbon structures. We performed a systematic study in which we showed the LB underestimates the fluid/solid interaction energies compared to the density functional theory (DFT) calculation results. To improve the classical description, we propose a new parametrization for the cross-terms of the Lenard-Jones (LJ) potential by fitting DFT forces and energies. The effects of the new parametrization on the gases adsorption within single-walled carbon nanotubes (SWCNTs) with varying diameters, are investigated with Grand Canonical Monte Carlo simulations. We observed considerable differences in the $CO_2$ and $CH_4$ density within SWCNTs compared to those obtained with the standard approach. Our study highlights the importance of going beyond the traditional LB mixing rules in studies involving solid/fluid interfaces of confined systems. The revised mixing terms enhanced fluid/carbon interface description with excellent transferability ranging from SWCNTs to graphene.

# 1    Introduction

The increasing interest in developing innovative technologies and processes for carbon capture and storage has driven several studies toward breakthroughs involving methane ($CH_4$) and carbon dioxide ($CO_2$) confined within carbon nanotubes (CNTS) [1]. CNTs emerged as a promising material for developing of new gas separation devices [2] due to their adsorption capability [3], fast mass transport [4], and mechanical strength [5,6].

Single-walled carbon nanotubes (SWCNTs) have been used as a host of a large variety of guest fillers, such as water [7], $CO_2$ [8], $CH_4$ [9], and hydrogen [10]. Computational molecular modeling is thus an essential tool for understanding underlying molecular mechanisms in chemical and physical processes of confined systems, providing atomic-level information that is often not accessible through experiments. However, the prediction success of these studies relies on using suitable force fields that provide an accurate description of the nanoscopic materials.

Previous studies also addressed the issue of $CO_2$ and $CH_4$ adsorption in graphene [11–13] and SWCNTs [14–16] within the density functional (DFT) approach. Other studies discussed how the available DFT exchange and correlation functionals describe these systems [12,13]. Using molecular dynamics (MD) and grand canonical Monte Carlo (GCMC) simulations, the adsorption of $CO_2$ and $CH_4$ molecules in carbon structures have been widely studied with a variety of force fields for both nanotube and gas molecules [8,9,17–20].

In general, MD and GCMC studies considered the Lorentz-Berthelot (LB) mixing rules for the cross parameters. For $CO_2$, it has been shown that different three-site models provide similar densities inside the CNT when the same potential is used for the SWCNTs [8]. The reason is that the Lennard-Jones parameters of these molecular models are usually similar. However, the available Lennard-Jones parameters for the SWCNTs may differ considerably, with energy potential well depths varying by almost 100%.

Employing the Steele potential [21] with LB mixing rules is a common approach in describing fluid/graphene-like interfaces [9,22–30]. Other studies [18,20,31] employed the same mixing rules but considered different CNT potentials, such as the AMBER96 [8,19]. These studies have not provided strong arguments towards the preference for a specific force field, neither have they compared the potentials with first-principles data or discussed their influence on the adsorption properties and fluid density, for instance.

The current work systematically compares several of the most popular existing force fields and provides optimized nonbonded cross parameters to describe $CO_2$–SWCNTs and $CH_4$–SWCNTs interfaces. These parameters were obtained by fitting interaction energies and forces from DFT calculations. The parameterization was tested in configurations out of the fitting set, displaying results very similar to the ones obtained with the DFT approach. Using GCMC simulations, we showed that the standard approach using the LB mixing rules underestimates the gas density inside the SWCNT, while our potential overcomes this limitation. Thus showing the importance of going beyond the traditional LB mixing rules for confined systems in the context of greenhouse gases adsorption in carbon nanostructures.

## 2    Methodology
### 2.1    DFT

Density functional theory (DFT) calculations were performed with the Siesta [32] package considering norm-conserving pseudopotentials, localized atomic orbitals with a double-zeta polarized (DZP) basis set, and 400 Ry mesh cutoff. Structure relaxations were performed considering the convergence criteria for the forces smaller than $10^{-4}$ eV/Å. Besides the standard local density functional (LDA) [33], three of the most popular exchange and correlation functionals that include van der Waals corrections (vdW-DFs) were employed in our calculations, namely, the functional by Klimes, Bowler, and Michaelides (KBM or optB86b-vdW) [34], the exchange-correlation potential parameterized by Cooper, known as C09 functional [35], and the BH exchange functional by Kristian Berland and Per Hyldgaard [36] (access the Supporting Information (SI) section S1 to see a brief description of these functionals).

The graphene sheet was modeled with a $2 \times 2 \times 1$ supercell with 30 Å vacuum to allow a similar molecular coverage compared to the experimental reference data [12]. We also considered a $5 \times 5 \times 1$ supercell configuration to exclude the interactions between neighboring molecules. For these calculations, we used a mesh equivalently to 16 x 16 x 1 k-points per unit cell within the Monkhorst-pack scheme to sample the Brillouin zone. Besides the graphene sheets, we performed a similar study on (8,8), (10,10), and (0,17) SWCNTs with 20 Å length to explore chirality and diameter effects.

## 2.2    Force Fields

Interaction energies were evaluated with the LAMMPS package [37] considering the implementation of non-bonded contribution within the classical force field given by:

$$U_{inter} = \sum_{i<j} 4\epsilon_{ij}\left[\left(\frac{\sigma_{ij}}{r_{ij}}\right)^{12} - \left(\frac{\sigma_{ij}}{r_{ij}}\right)^{6}\right] + \sum_{i<j} \frac{q_i q_j}{r_{ij}}, \qquad (1)$$

where $\epsilon_{ij}$, $\sigma_{ij}$ are the Lennard-Jones parameters, $q_{i(j)}$ the atomic charge, and $r_{ij}$ is the separation distance. The molecules and carbon structures were kept rigid during the simulations.

We considered six different sets of Lennard-Jones parameters for the graphene/CNT system, taken from AIREBO [38], Mao [39], Huang [18], AMBER96 [40], Walther [41], and Steele [21] force fields, all previously employed in adsorption studies. For $CH_4$, we considered the OPLS-AA [42] potential, while $CO_2$ was modeled using the EPM2 [43] model, these being among the most popular and accurate force fields found in the literature. All the parameters are given in Table S1 in SI. For these force fields, cross terms describing the interaction between atomic species were obtained with the Lorentz-Berthelot mixing rules, i.e., $\epsilon_{ij} = \sqrt{\epsilon_{ii}\epsilon_{jj}}$ and $\sigma_{ij} = (\sigma_{ii} + \sigma_{jj})/2$, as it was used in several works found in the literature [9, 22-30].

Graphene-like structures potentially have three adsorption sites, i.e., top (T), bridge (B), and hollow (H), as depicted in Figure 1a. Among the possibilities, some typical and stable molecular orientations besides these sites (see Figures 1a-g) were considered in our calculations to evaluate the interaction energy curves and forces with both DFT and classical FFs. Within these approaches, the interaction energy, $E_{int}$, is given by

$$E_{int} = E_{tot} - [E_{Car} + E_{Mol}], \qquad (2)$$

where $E_{tot}$ is the system's total energy; $E_{Car}$ and $E_{Mol}$ are the energies of the isolated carbon material (CNT or graphene) and the adsorbed molecule, respectively.

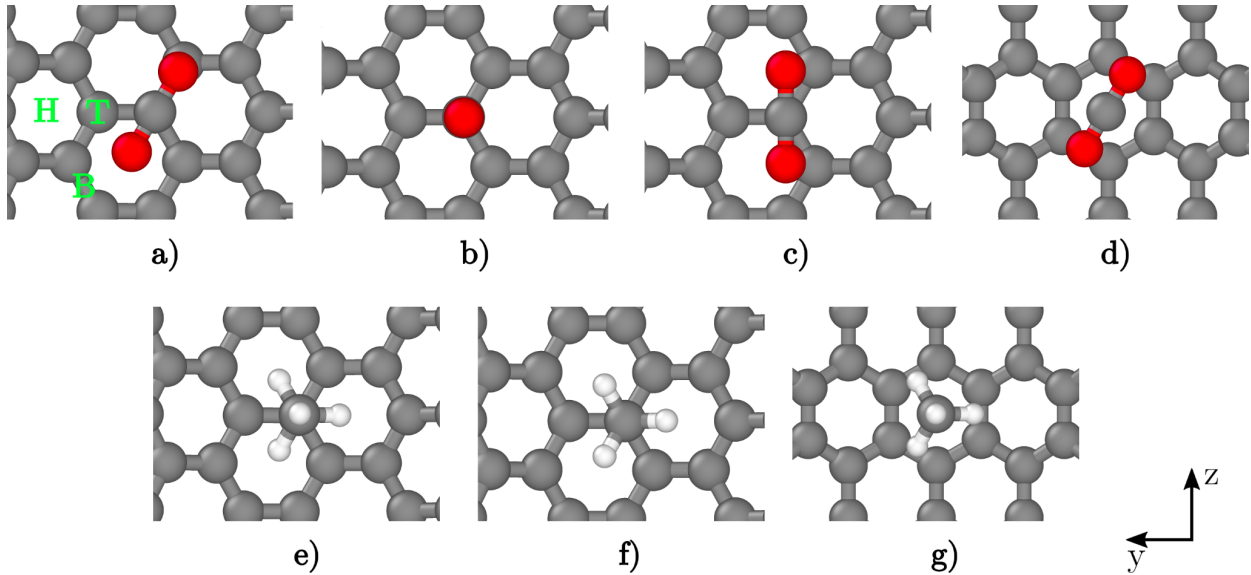

Figure 1: Some common and stable molecular configurations on the carbon structures's adsorption sites for both $CO_2$ a-d and $CH_4$ e-g considered in our calculations. Colors: Gray (carbon), red (oxygen), white (hydrogen). Figure 1a also exhibits the typical adsorption sites – hollow (H), top (T), bridge (B) – in graphene-like lattices.

## 2.3 Parametrization process

We obtained a new set of Lennard-Jones cross parameters optimized for the $CO_2$-SWCNT and $CH_4$-SWCNT interfaces using the DFT interaction energies and atomic forces. The data was fitted with the GULP package [44]. To obtain a better fit around the minimum of the energy curve, and provide a better description of the equilibrium distances, we weighted ($w$) the configurations according to the equation:

$$w = \exp[(E_{int} - E_{int}^{min})/kT], \qquad (3)$$

where $E_{int}^{min}$ is the minimum energy point among the curves used in the fitting process and temperature $T$ was defined as 200 K and 300 K for $CH_4$ and $CO_2$, respectively. These values were selected to obtain a good balance between the description of the interaction energy minimum and over the entire curve. The weight for the interaction energy was multiplied by $10^3$ to get similar contributions as for the forces. This procedure is necessary because there are $3N$ (with $N \approx 350$ being the number of atoms in each calculation) force components for each interaction energy point; therefore, the $10^3$ factor evens the contribution between the forces and energy during the fitting.

## 2.4 GCMC

To evaluate the impact of the force fields on the adsorption of the gases within SWCNTs, we performed GCMC simulations using the Cassandra package [45]. With this approach, first the chemical potentials of both $CO_2$ and $CH_4$ at 300 K and 1 atm were obtained using the Widom insertion method [46]. These simulations were performed in the *NPT* ensemble with $1 \times 10^6$ followed by $3 \times 10^6$ steps in thermalization and production phases, respectively.

Subsequently, the SWCNTs with 200 Å length were loaded with up to 200 gas molecules, depending on their diameter. An *NVT* simulation was run for $5 \times 10^4$ steps to generate the initial configuration. This process was followed by the GCMC using insertions restricted by the SWCNTs cylinder geometry. These simulations were performed for $5 \times 10^6$ steps, using configurational-biased insertions (with 16 trial insertions) and considering probabilities of 25% for each of the translation, rotation, insertion, and deletion moves. The simulations were long enough to reach the convergence of both internal energy and the number of adsorbed molecules.

## 3 Results and Discussions

### 3.1 DFT functional

We benchmarked the DFT functionals against experimental data to establish the most appropriate functional to describe the interaction between $CO_2$ and $CH_4$ molecules and the graphene monolayer. The interaction energy curves obtained by applying Equation 2 as a function of distance $r$ from the 2x2x1 graphene sheet are shown in Figure **2a** ($CO_2$) and Figure **2b** ($CH_4$). Comparing these curves for $CO_2$, BH gives a slight deviation from C09 and KBM functional. The equilibrium distance and energy predicted by BH are 3.20 Å and -5.83 kcal/mol, respectively. These values for the C09 are 3.11 Å and -6.44 kcal/mol; while the KBM gave 3.12 Å and -6.44 kcal/mol. Our results for the C09 and KBM are in good agreement with the experimental results reported in the literature (-6.26 kcal/mol) under similar conditions [12]. LDA displayed a significant deviation despite the equilibrium distance being in good agreement with C09 and KBM, showing the importance of considering the long-range contributions in the DFT calculations.

Osouleddini et al. [11], using the hybrid X3LYP functional, reported the $CO_2$-graphene adsorption energy and equilibrium distance as being -0.69 kcal/mol and 3.12 Å, respectively. In turn, Takeuchi et al. [12] using vdW-DF1, optB86b-vdW (KBM), and rev-vdW-DF2, found $E_{int}$ equal to -5.80, -5.90, and -4.08 kcal/mol, respectively. They also reported the equilibrium distance for the vdW-DF1 and optB86b-vdW as being 3.4 Å and 3.2 Å, respectively. The minor differences observed in their KBM result in comparison to ours may be attributed to the use of different geometries, software, and a distinct basis set. Other $E_{int}$ values reported in the literature range from -4.11 to -5.99 kcal/mol [47].

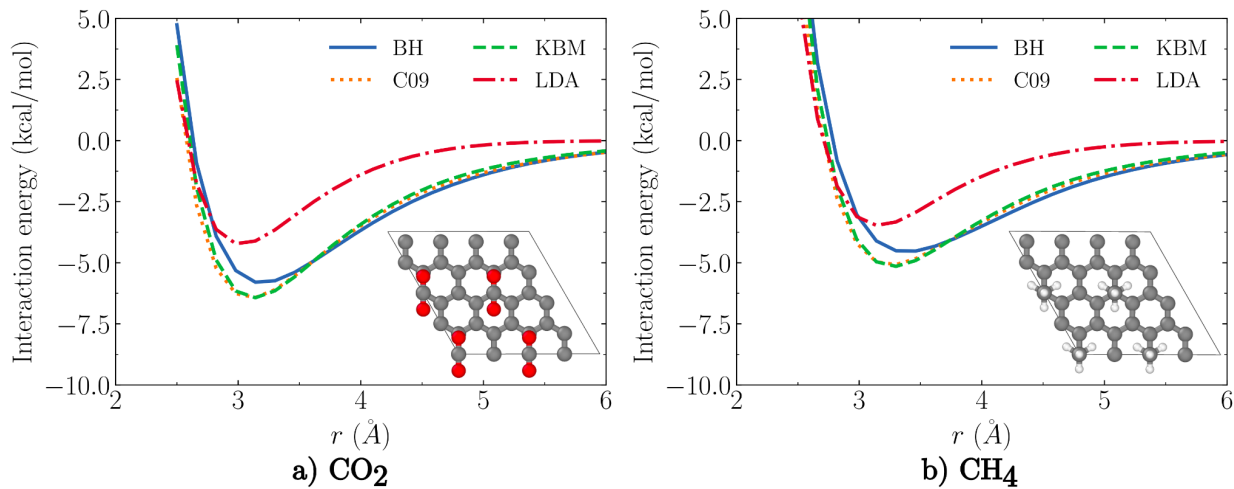

**Figure 2**: The interaction energy curves of $CO_2$-graphene (a) and $CH_4$-graphene (b) (2x2x1 supercell) interface using the LDA, BH, KBM, and C09 functionals. For this study, we considered the most energetically favored adsorption configuration represented in Figure 1a and Figure 1e, for $CO_2$ and $CH_4$, respectively.

The $CH_4$-graphene interaction energy curves display a similar pattern observed for the $CO_2$ case, (see Figure 2b). For the BH functional, the obtained equilibrium distance and minimum interaction energy were 3.39 Å and -4.55 kcal/mol, respectively. For C09, these values are 3.26 Å and -5.09 kcal/mol, and for KBM, 3.28 Å and -5.15 kcal/mol. For the same configuration, Thierfelder et al. [13] found the equilibrium distance ranging from 3.05 to 3.99 Å and minimum interaction energy varying from -0.23 to -7.38 kcal/mol, using several DFT functional flavors and second order Møller–Plesset perturbation theory. In contrast, Osouleddini et al. [11] found a similar equilibrium distance of 3.35 Å, but lower adsorption energy of -1.15 kcal/mol.

Most of the experimental results are available for the methane-graphite whereby the equilibrium distance and desorption energy ranges from 3.21 to 4.27 Å and 2.73 to 4.6 kcal/mol, respectively [48]. In the absence of accurate experimental results for similar coverage of the $CH_4$-graphene system, such as the ones available for $CO_2$ [12], we argue that any of the studied van der Waals functionals can describe the interfaces accurately enough for the fitting with Lennard-Jones parameters, since our results are in the same range as the values reported in the literature for the methane-graphite interface.

There are no strong arguments towards selecting either C09 or KBM, as both give very similar results, compatible with the experimental desorption energies for $CO_2$. As an advantage, the KBM functional has been implemented so far in more *ab initio* software than C09. Comparing C09 and KBM implementations in Siesta, the KBM performed more efficiently than C09 for our systems. For these reasons, we chose the KBM functional to extract the new set of Lennard-Jones cross parameters optimized for the $CO_2$-SWCNT and $CH_4$-SWCNT description for consistency.

## 3.2  FF parametrization

We evaluated the performance of the selected classical FFs using the Lorentz-Berthelot mixing rules in describing the $CO_2$-SWCNT and $CH_4$-SWCNT interfaces, by comparing the corresponding interaction energy curves with the DFT results (see section S3 for more details). For the investigated adsorption configurations, all FFs curves exhibited a smaller depth than the DFT-KBM functional (see Figures 3 a-b). Also, the equilibrium distance is slightly smaller, with the Walther potential displaying the greatest discrepancy. These differences can lead to an inadequate description of the gas adsorption on the carbon structures.

The inconsistency observed between classical FFs and DFT calculations suggests the need for a more refined parametrization. By using our approach to fit the Lennard-Jones cross parameters, $\epsilon_{ij}$ and $\sigma_{ij}$, to the DFT-KBM dataset, we obtained the optimized parameters shown in **Table 1**. More details about the dataset are given in section S3.

The energy curve with the optimized parameters agrees with the DFT results and improves the forces, potential well depth, and equilibrium distances concerning the standard approach. The quality of our fit can be illustrated by the correlation between the classical and DFT-KBM interaction energies. We obtained a Pearson correlation coefficient for the fitted and DFT interaction energies of 0.953 and 0.974 for $CO_2$ and $CH_4$, respectively. The forces' correlations are smaller, having coefficients of up to 0.699 and 0.659 for $CO_2$ and $CH_4$, respectively. Figures showing the correlations and further discussion are presented in section S4.

**Table 1:** Crossing terms for the LJ potential optimized by the fitting process.

|  | $\epsilon_{ij}$ (K) | $\sigma_{ij}$ (Å) |
|---|---|---|
| $C_{CNT}$-$C_{CO2}$ | 74.466 | 3.09 |
| $C_{CNT}$-$O_{CO2}$ | 85.189 | 3.09 |
| $C_{CNT}$-$C_{CH4}$ | 66.540 | 3.38 |
| $C_{CNT}$-$H_{CH4}$ | 49.609 | 2.57 |

Figure 3 a-b) illustrates the comparison between the interaction energies computed with the FFs, our parameterized cross terms, and the DFT-KBM models. We highlight that one of the most used FFs for describing adsorption in such systems, Steele's FF, shows the smallest $E_{int}$, in disagreement with the DFT data. Our fitted parameters also improve the energy difference at the center of the nanotube and the minimum. This value is important for adsorption studies as it is what is effectively seen by the Monte Carlo simulations.

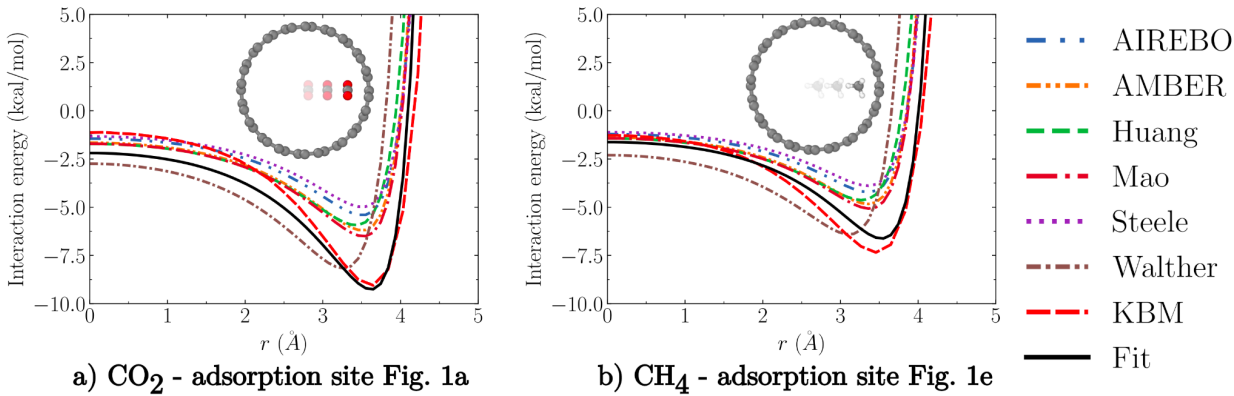

a) $CO_2$ - adsorption site Fig. 1a     b) $CH_4$ - adsorption site Fig. 1e

**Figure 3:** The interaction energy curves of a) $CO_2$ and b) $CH_4$ adsorbed on the (10,10) SWCNT. Here we compare the classical FFs performance against the DFT (KBM), and the parameterized potential (Fit).

We tested the obtained parameterization in different configurations not present in the fitting set (see Figures 4a-d and section S3 of SI). By considering distinguished systems, we evaluated the transferability of the LJ parameters to other chemical environments. In these new systems, the fitted curve showed a remarkable agreement with KBM results, even though these configurations were not considered in the fitting process.

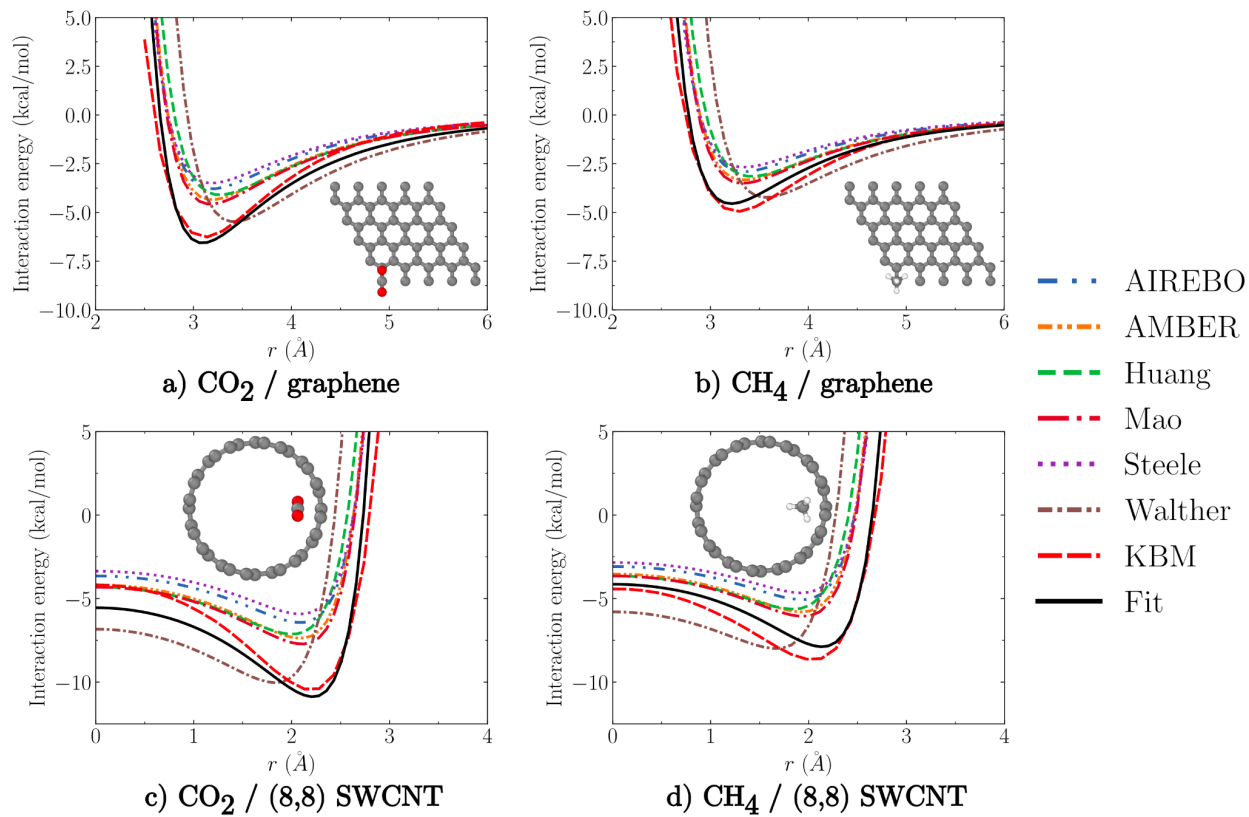

**Figure 4**: The interaction energy curves of a) $CO_2$ and b) $CH_4$ above the graphene sheet and c) $CO_2$ and d) $CH_4$ adsorbed in an (8,8) SWCNT. These systems were apart from the fitting set configurations, therefore, showing how the fitted parameters perform for interactions beyond the fitted situations.

In all the cases analyzed, our fitted model improved the adsorption energy description with respect to the existing FFs combined with the traditional Lorentz-Berthelot mixing rules, confirming the parametrization accuracy in describing fluid/-carbon interfaces. Also, the fitted and KBM curves agreed well with the energy depth and equilibrium distance for both a smaller radius SWCNT and the infinite limit (graphene), showing their remarkable transferability capability.

### 3.3 Gas adsorption in the SWCNTs

The discrepancy observed between classical FFs employing Lorentz-Berthelot mixing rules and the DFT calculations indicates that the structuring and distribution of gas molecules inside the SWCNTs may also differ. In this context, we compared the existing classical FFs and that with optimized LJ cross parameters in describing these properties as a function of the tube diameter. Figures **5a-b** show the gas densities obtained with GCMC as a function of tube diameter in (6,6), (12,12), and (16,16) SWCNTs.

The fluid density within the CNTs strongly depends on the adopted force field and tube diameter. For the $CO_2$ case, we observed the AIREBO and Steele curves give a consistent decrease in fluid density as the nanotube diameter increases (see Figure **5a**). Regarding the other FFs, the density increases as the nanotube increases from 8.1 Å (6,6) to 16.3 Å (12,12), and decreases as the nanotube increases from 16.3 Å (12,12) to 21.7 Å (16,16). A similar trend was observed by Alexiadis and Kassinos [8] using MD. Also, the amount of $CO_2$ in the nanotubes predicted by the traditional classical FFs is much lower than that obtained with our fitted model. The densities obtained within the Steele potential and LB mixing rules can be up to seven times smaller than the ones predicted by our fitted potential for the (16,16) SWCNT.

For the $CH_4$ case (Figure **5b**), density curves obtained with AIREBO, Mao, Huang, AMBER, and Steele curves display a similar trend, with the densities decreasing as the nanotube diameter increases from 8.1 Å (6,6) to 16.3 Å (12,12) and remains practically unchanged from 16.3 Å to 21.7 Å (16,16). These results differ from those obtained with the Walther potential and our fitted parameters, since the $CH_4$ density slightly increases as the nanotube increases from 8.1 Å to 16.3 Å and then decreases as the nanotube increases from 16.3 Å to 21.7 Å. However, the densities in the (12,12) and (16,16) are almost 50% greater than Walther's using the fitted parameters. We reach the same conclusion as for the $CO_2$ case, i.e., the number of adsorbed molecules is underestimated by all the force fields tested. These results revealed the importance of revisiting the adsorption of greenhouse gases in carbon structures to approach the DFT achievements.

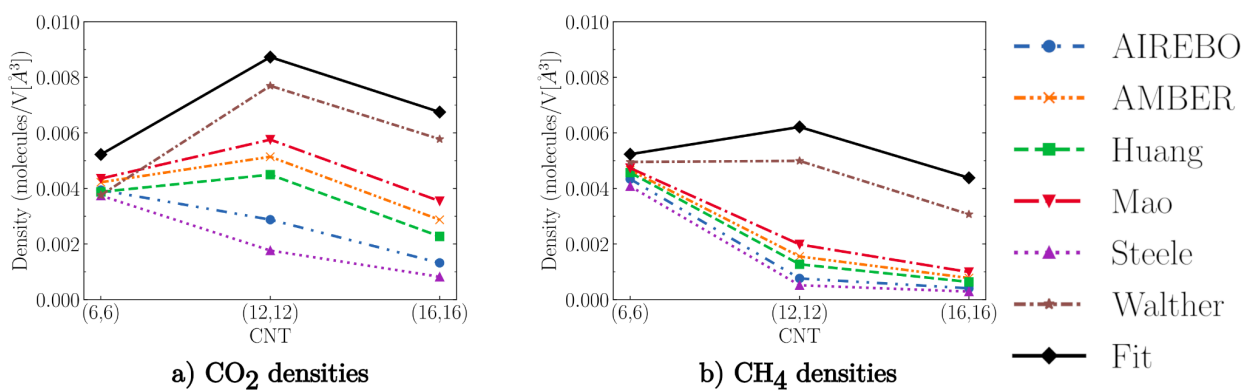

a) $CO_2$ densities          b) $CH_4$ densities

**Figure 5:** Comparison of the density of a) $CO_2$ and b) $CH_4$ in (6,6), (12,12), and (16,16) CNTs.

## 4    Conclusions

The differences observed in the performance of DFT functionals and classical FFs in describing the adsorption of $CH_4$ and $CO_2$ in carbon structures have been addressed in our investigation. We proposed a new set of LJ cross parameters obtained by fitting DFT energies and forces. The accuracy in estimating the adsorption properties has been established by a systematic comparison with DFT results.

Among the DFT functionals considered in our study, the KBM flavor could properly describe the adsorption of $CO_2$ and $CH_4$ in carbon structures in accordance with previous experimental studies. Considering different adsorption sites and nanotube chirality, we have fitted a new set of Lennard-Jones cross parameters $\epsilon_{ij}$ and $\sigma_{ij}$. These parameters enhanced the fluid/solid interface description considerably with respect to the existing FFs, which use the standard Lorentz-Berthelot mixing rules. Most FFs underestimate the adsorption energy and equilibrium distances. Regarding the adsorption energies, the Walther potential has a deeper well if compared with the other FFs, but it is closer to the reference DFT calculations, being about 20% shallower. These differences with respect to the DFT results can be up to 100% for the other FFs, including the popular Steele and AMBER potentials.

The fitted model was employed beyond the fitting set, and it showed great transferability capability. Furthermore, using Grand Canonical Monte Carlo (GCMC), we investigated the adsorption features of $CO_2$ and $CH_4$ in SWCNTs of different sizes. Our GCMC simulations showed that the standard approach using the Lorentz-Berthelot mixing rules underestimates the gas density inside the SWCNT. These results highlight the importance of going beyond the Lorentz-Berthelot mixing rules involving fluid/carbon interfaces. We highlight that our findings may improve the molecular modeling studies involving challenging problems within the carbon dioxide capture, utilization and storage context.

## Acknowledgements

We gratefully acknowledge support of the RCGI – Research Centre for Gas Innovation, hosted by the University of São Paulo (USP) and sponsored by FAPESP – São Paulo Research Foundation (2014/50279-4 and 2020/15230-5) and Shell Brasil, and the strategic importance of the support given by ANP (Brazil's National Oil, Natural Gas and Biofuels Agency) through the R&D levy regulation. We also thank the financial support of FAPESP through project numbers #2017/02317-2, #2019/21430-0, and #2020/01558-9, and the National Council for Scientific and Technological Development (CNPq) through grant 307064/2019-0 for financial support. HMC also thanks the Research Council of Norway through the Centre of Excellence Hylleraas Centre for Quantum Molecular Sciences (grant number 262695). The computational time for the calculations was provided by High-Performance Computing facilities at the University of São Paulo (USP). We thank Julian D. Gale for the insightful discussions regarding the parametrization using GULP.

**CRediT authorship contribution statement**

**HMC:** Conceptualization, Formal analysis, Investigation (classical simulations), Software, Visualization

**TDL:** Methodology, Software, Formal analysis, Investigation (DFT), Visualization

**DAD:** Conceptualization and Writing - original draft

**AK:** Formal analysis, Writing - review & editing

**CRM:** Conceptualization, Resources, Writing - review & editing, Supervision, Funding acquisition

# Supporting Information: Revisiting greenhouse gases adsorption in carbon nanostructures: advances through a combined first-principles and molecular simulation approach

**Authors:** Henrique M. Cezar, Teresa D. Lanna, Daniela A. Damasceno, Alexsandro Kirch, Caetano R. Miranda
Universidade de Sao Paulo, Instituto de Fisica, Rua do Matao 1371, Sao Paulo, SP 05508-090, Brazil
Corresponding author: Caetano R. Miranda, email: crmiranda@usp.br

## S1 Exchange and correlation functionals

The optB86b exchange functional, developed by Klimes, Bowler, and Michaelides, also known as KBM or optB86b-vdW [S1], is the optimized Becke86b [S2] exchange combined with the vdW-DF1 correlation [S3]. This functional leads to a smaller error cancellation between exchange and the overestimated correlation than the optB88-vdW functional. The exchange-correlation potential parameterized by Valentino, known as C09 functional [S4], demonstrates remarkable improvements in intermolecular separation distances compared with previous functionals, while improving the accuracy of vdW-DF interaction energies [S4]. The BH [S5] exchange functional, designed by Kristian Berland and Per Hyldgaard, tests the robustness of the plasmon description vdW-DF, which gives a good description of both exchange and correlation effects in the low-to-moderate regime.

## S2 Classical Force Fields

Table S1: Force field parameters of the selected force fields considered in our study.

| | $CO_2$ (EPM2) | | $CH_4$ (OPLS-AA) | | SWCNT | | | | | |
|---|---|---|---|---|---|---|---|---|---|---|
| | C | O | C | H | AIREBO | AMBER 96 | Mao | Huang | Steele | Walther |
| $\epsilon_{ii}$ (K) | 28.129 | 80.507 | 33.212 | 15.097 | 32.96 | 43.30 | 48.78 | 35.26 | 28.00 | 52.87 |
| $\sigma_{ii}$ (Å) | 2.757 | 3.033 | 3.500 | 2.500 | 3.400 | 3.400 | 3.370 | 3.550 | 3.400 | 3.851 |
| $q_i$ (e) | +0.6512 | -0.32560 | -0.24 | +0.06 | 0.0 | 0.0 | 0.0 | 0.0 | 0.0 | 0.0 |

## S3    Fitting procedure and additional results

The fitting process was carried out using the GULP package trying to match both energies and forces from single-point DFT calculations performed with the KBM exchange-correlation functional. The calculations were performed considering the initial geometry obtained with the classical models employed for $CO_2$ and $CH_4$ and unrelaxed geometry for the SWCNTs generated with  the visual molecular dynamics (VMD) [S6]. To remove the extra force due to the system not being relaxed, we subtracted the forces of the isolated systems from the total forces. This subtraction allows removing some of the forces related to the geometry, while also improving the forces by removing some spurious forces resulting from the eggbox effect present in calculations performed with Siesta.

In Figures S1 and S2, we show the fitted interaction energies, for $CO_2$ and $CH_4$, respectively, together with the curves obtained using the classical force fields mentioned in the main text. The curves shown in the figures represent the complete dataset used for the fitting. As presented in the main text, the fitted parameters were tested on configurations other than the ones shown in Figures S1 and S2 and exhibited great transferability (see Figure 4 in the main text). Figures S3 a) and b) illustrate a scatter plot of the energies showing the correlation between the fitted and DFT values for $CO_2$ and $CH_4$, respectively. The Pearson correlation coefficients were 0.953 for $CO_2$ and 0.974 for $CH_4$.

Since we also considered the forces in the fitting process, we compared the differences between the force components obtained with the fitted force field and with DFT, as shown in Figure S4. As observable, most of the force differences lie between -0.05 and 0.05 eV/Å. Even though there were force differences greater than those displayed in the interval of plus or minus 0.15 eV/Å, the frequencies were so small that the histogram bars were not visible at this scale. These outliers can be better seen in the scatter plot showing the correlations between the fitted force field force components and DFT force components, depicted in Figure S5.

For $CO_2$ (Figure S5 a) only two points, corresponding to the adsorption site shown in Figure 1 b) of the main text, show a larger deviation from the baseline correlation depicted by the diagonal dashed line. This adsorption site is not favorable in comparison with the others, as shown in Figure S1 c). Even considering all the points, the Pearson correlation for the forces components of $CO_2$ is 0.699.

The situation is similar for $CH_4$. Even though the picture shown in Figure S5 b exhibits a line of points in which the fitted forces are close to zero while having considerably different values with DFT, these points correspond to only 0.26 % of the total number of forces. These components correspond to the configuration shown in Figure 1 f), where one of the hydrogens approaches the SWCNT on the top site. This adsorption site shows considerably higher interaction energies, as shown in Figure S2 b). If these components are removed from the picture, as shown in Figure S6, we observe a Pearson correlation of 0.659.

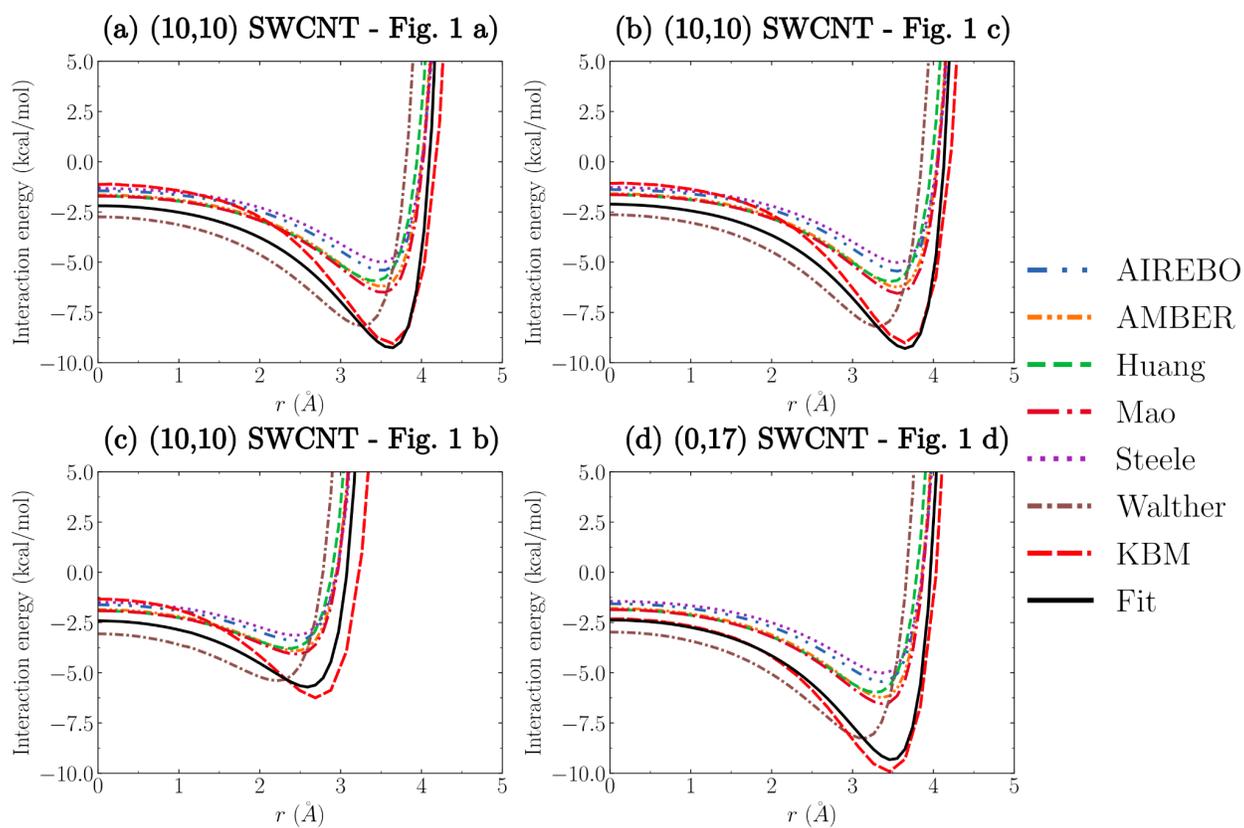

**Figure S1:** The interaction energy curves of $CO_2$-SWCNTs using the AIREBO, Mao, Huang, AMBER, Walther, and Steele force fields, KBM functional and Fit. The SWCNT and adsorption sites (in reference to Figure 1 of the main text). The DFT results in a-d) figures were used in the fitting process that resulted in the parameters used to generate the black solid line curve.

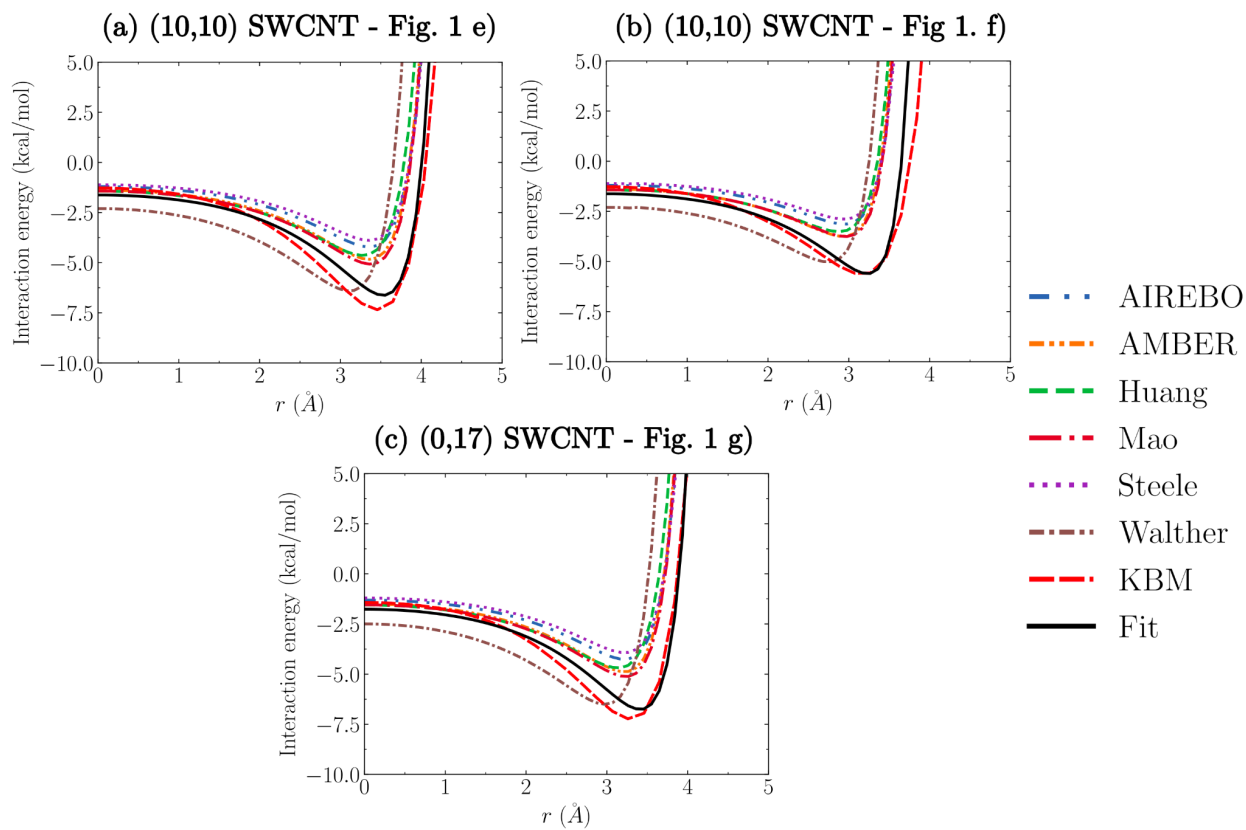

**Figure S2:** The interaction energy curves of CH$_4$-SWCNTs using the AIREBO, Mao, Huang, AMBER, Walther, and Steele force fields, KBM functional and Fit. The SWCNT and adsorption sites (in reference to Figure 1 of the main text). The DFT results in a-c) figures were used in the fitting process that resulted in the parameters used to generate the black solid line curve.

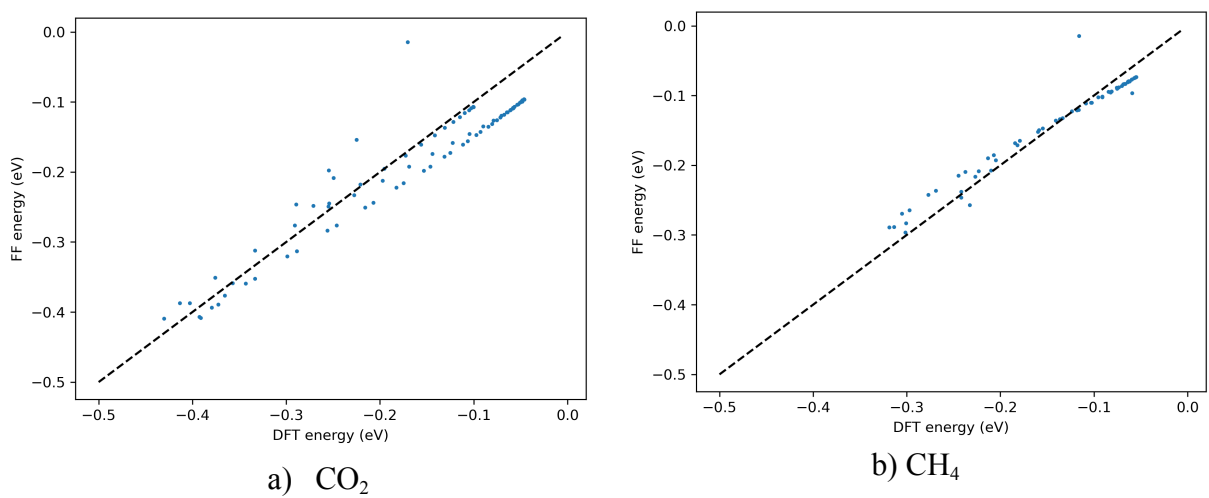

a) CO$_2$

b) CH$_4$

**Figure S3:** Scatter plot of the interaction energies obtained with DFT and the fitted potentials for a) $CO_2$ and b) $CH_4$. The diagonal dashed line represents where perfectly correlated points should lie.

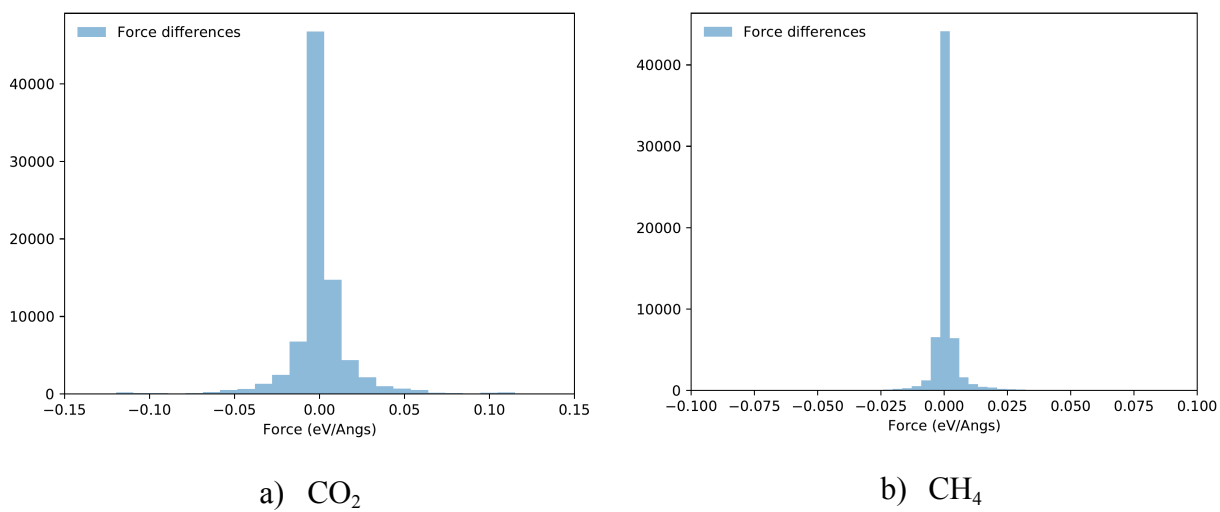

a)  $CO_2$  b)  $CH_4$

**Figure S4:** Distribution of force differences ($F_{DFT} - F_{Fit}$) for each force component for a) $CO_2$ and b) $CH_4$.

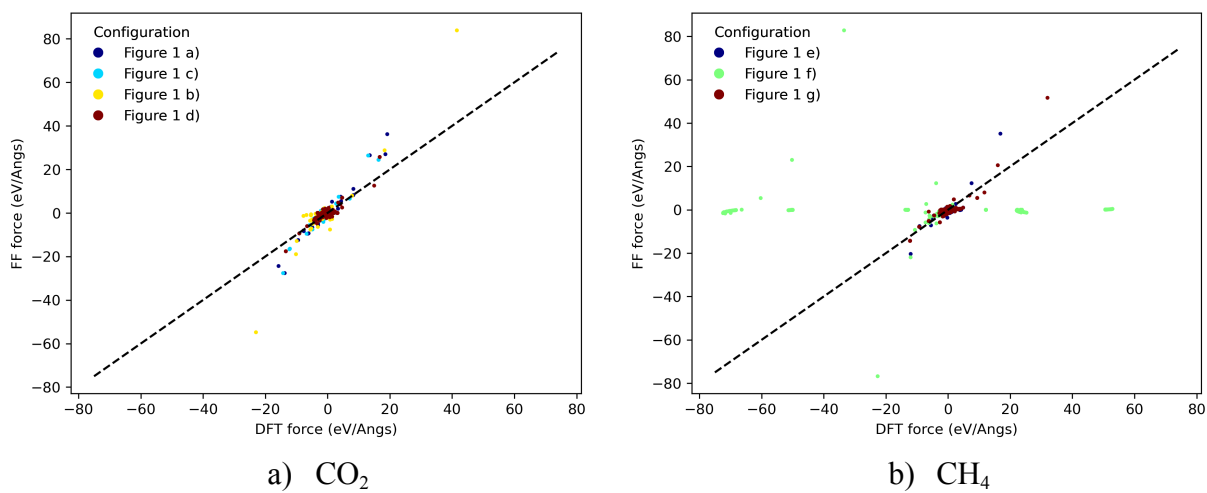

a)  $CO_2$  b)  $CH_4$

**Figure S5:** Scatter plot of the force components obtained with DFT and the fitted potentials for a) $CO_2$ and b) $CH_4$. The diagonal dashed line represents where perfectly correlated points should lie. This figure includes forces for all the atoms in the system for all the trajectories included in the fitting set. The color coding references the adsorption sites in Figure 1 of the main text. For methane (right panel, b) the green points correspond to the configuration in which one of the hydrogens approaches the CNT on a top site, a configuration with higher energy in comparison with the other adsorption sites.

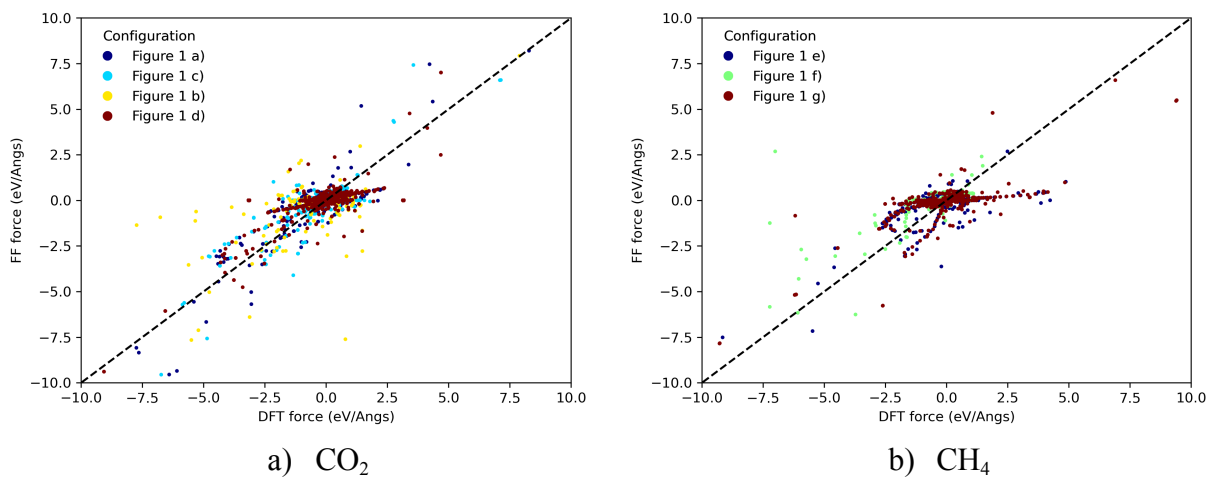

a) $CO_2$   b) $CH_4$

**Figure S6:** Scatter plot of the force components obtained with DFT and the fitted potentials for a) $CO_2$ and b) $CH_4$. The diagonal dashed line represents where perfectly correlated points should lie. This figure includes force components with absolute value smaller than 10 eV/Å. The color coding references the adsorption sites in Figure 1 of the main text.